\documentclass[twocolumn,aps,superscriptaddress]{revtex4}
\usepackage{amsmath,bm}
\usepackage{amssymb}
\usepackage{graphicx}
\usepackage[normalem]{ulem}
\usepackage[dvips]{color}
\usepackage{appendix}
\usepackage{dcolumn}
\usepackage[colorlinks,
           linkcolor=blue,
            anchorcolor=black,
            citecolor=blue
            ]{hyperref}
\usepackage[all]{hypcap}

\makeatletter

\newcommand{\Rmnum}[1]{\expandafter\@slowromancap\romannumeral #1@}
\makeatother

\begin{document}
\title{Proton correlations and apparent intermittency in the UrQMD model with hadronic potentials}

\author{Pengcheng Li}
\affiliation{School of Nuclear Science and Technology, Lanzhou University, Lanzhou 730000, China}
\affiliation{School of Science, Huzhou University, Huzhou 313000, China}

\author{Yongjia Wang}
\affiliation{School of Science, Huzhou University, Huzhou 313000, China}

\author{Jan Steinheimer}
\affiliation{Frankfurt Institute for Advanced Studies, Ruth-Moufang-Str. 1, D-60438 Frankfurt am Main, Germany}

\author{Qingfeng Li\footnote{Corresponding author: liqf@zjhu.edu.cn}}
\affiliation{School of Science, Huzhou University, Huzhou 313000, China}
\affiliation{Institute of Modern Physics, Chinese Academy of Sciences, Lanzhou 730000, China}

\author{Hongfei Zhang}
\affiliation{School of Nuclear Science and Technology, Lanzhou University, Lanzhou 730000, China}
\affiliation{Joint Department for Nuclear Physics, Lanzhou University and Institute of Modern Physics, Chinese Academy of Sciences, Lanzhou 730000, China}
\date{\today}

\begin{abstract}
It is shown that the inclusion of hadronic interactions, and in particular nuclear potentials, in simulations of heavy ion collisions at the SPS energy range can lead to obvious correlations of protons. These correlations contribute significantly to an intermittency analysis as performed at the NA61 experiment. The beam energy and system size dependence is studied by comparing the resulting intermittency index for heavy ion collisions of different nuclei at beam energies of $40A$, $80A$ and $150A$ GeV. The resulting intermittency index from our simulations is similar to the reported values of the NA61 collaboration, if nuclear interactions are included. The observed apparent intermittency signal is the result of the correlated proton pairs with small relative transverse momentum $\Delta p_{t}$, which would be enhanced by hadronic potentials, and this correlation between the protons is slightly influenced by the coalescence parameters and the relative invariant four-momentum $q_{inv}$ cut.
\end{abstract}

\maketitle

\section{INTRODUCTION}
The exploration of the properties of the Quark Gluon Plasma (QGP) \cite{QGP1,QGP2,QGP3} and the phase structure of quantum chromodynamics (QCD) \cite{QCD1973} are the main objectives in relativistic heavy-ion collisions (HICs). Lattice QCD calculations at vanishing baryon chemical potential ($\mu_B$) determined a chiral crossover transition at $T\simeq155$ MeV \cite{PRD71034504,Lattice,PRD90094503}. Various theoretical investigations suggest that at finite baryon chemical potential and temperature, there could be a critical end point (CEP) in the QCD phase diagram \cite{1stptcep1,1stptcep2,1stptcep3,1stptcep4}. In order to explore the QCD phase diagram at finite net-baryon density, one tries to vary the temperature and baryon chemical potential of the nuclear matter created in HICs by changing the colliding energy and size of the colliding nuclei as well as the centrality, leading to different freeze-out conditions in $\mu_B$ and $T$ \cite{whitepaper,SPSC-P-330-ADD-4,SCAN1}.

A non-monotonic behavior of fluctuation of observables is expected as a signal for the CEP \cite{PRL835435,PRL852072,PRL95182301,PLB633275,PRL102032301,sspma011012}. To investigate the phase transition and search for the CEP in HICs, a lot of analyses on fluctuations were done. For example, the analysis of event-by-event fluctuations of integrated quantities like the net-baryon number, electric charge, and strangeness, in particular their higher order cumulants \cite{NST281122017,prd96074510, prd101074502,prc101034915,pp583,prl126092301}. In addition, the analysis of local power-law fluctuations of the net-baryon density \cite{NA49SFM,Local} has drawn much attention. It can be detected through the measurement of the scaled factorial moments (SFMs) in transverse momentum space within the framework of a proton intermittency analysis \cite{SFM1,SFM2,Antoniou2005,Antoniou2006}. During the last decade, NA49 and NA61 performed a systematic search for critical fluctuations utilizing an intermittency analysis in central A+A collisions \cite{NA49prc81,Davis2002}. An indication of power-law fluctuations, in the transverse momentum phase space of protons at mid-rapidity in the central (0-12.5\% ) Si+Si collisions at 158$A$ GeV/$c$, has recently been presented \cite{NA49SFM}. In addition, a non-trivial intermittency effect was shown in a preliminary analysis of $^{40}$Ar+$^{45}$Sc collisions at 150$A$ GeV/$c$ \cite{NA61,Davisapb}.

A more indirect interpretation of recent STAR data also suggested a non-trivial intermittency index which cannot be described by a pure cascade simulation of the Ultra-relativistic
Quantum Molecular Dynamics (UrQMD) model, and therefore warrants further investigation  \cite{plb801135186,appburqmd}. In this work, effects of hadronic potentials on the intermittency behaviour in HICs will be evaluated by a systematic scan in collision energy and system size within a modified version of the UrQMD model with and without hadronic potentials.

This paper is organized as follows: in section \ref{theory}, we briefly present the scaled factorial moments and the transport model (UrQMD model) used in this work. In section \ref{results}, the result for the intermittency effect is discussed. Finally in section \ref{summary}, the results and conclusions will be summarized.

\section{Methodology}
\label{theory}
In order to investigate the non-statistical fluctuations and quantitatively understand the intermittent behavior in interactions, a technique based on the scaled factorial moments (SFMs) measurement was first introduced by Bialas and Peschanski \cite{SFM1,SFM2}. Here, momentum space is partitioned into equal-size bins and the SFMs are defined as:
\begin{equation}
  F_{q}(M)=\frac{\left\langle\frac{1}{M^D}\sum_{i=1}^{M^D}n_{i}(n_i-1)...(n_i-q+1)\right\rangle}{\left\langle\frac{1}{M^D}\sum_{i=1}^{M^D}n_i\right\rangle^{q}},
 \label{Eq:FM}
\end{equation}
where $M^D$ is the number of equally sized bins in which the D-dimensional space is partitioned, $n_i$ is the number of particles in the $i$-th bin, the angular brackets denote an average over bins and events and $q$ is the rank of the moment.

For a critical system, the fluctuations of the order parameter are self-similar \cite{Vicsek}, and the SFMs scale with the number of bins. In other words, if the dynamical fluctuations are self-similar in nature in the limit of small bin size, the SFMs are expected to increase with bin size following a power law. This effect is called intermittency, and can be quantified by this dependence:
\begin{equation}
F_{q}(M)\sim (M^{D})^{\phi_{q}},~~~~M\rightarrow\infty.
 \label{Eq:PowerLaw}
\end{equation}
The exponent $\phi_{q}$ is the so called intermittency index, which not only characterizes the strength of the intermittency, but also correlates with to the anomalous fractal dimension $d_{q}$ of the system \cite{Antoniou2006}. Many investigations suggest that one can explore the possible critical region of the QCD matter by using an intermittency analysis \cite{NA49SFM,Antoniou2006,NA49prc81,plb801135186,CriticalReg}. Universality class arguments associate the power-law behaviour of the SFMs with a specific exponent $\phi_{2}=\frac{2}{3}$ in case of the chiral condensate as order parameter \cite{Antoniou2005} and $\phi_{2}=\frac{5}{6}$ for the net-baryon density as order parameter \cite{Antoniou2006,NA49SFM}.

Since such an analysis of higher order SFMs requires a large amount of data one usually only considers the second scaled factorial moment (SSFMs) in transverse momentum space, which is obtained by setting $q=2$ and $D=2$ in Eq.~\ref{Eq:FM}.

The above considerations are true for idealized systems which can be studied in the thermal limit. In fact, heavy-ion collisions will never be such an ideal system for a number of reasons
\cite{NA49SFM}. To reveal the predicted power-law exponents, it is necessary to consider a large number of background effects. Such effects include the finite lifetime of the system, global and local conservation laws, finite experimental acceptance, freeze out dynamics, resonance decays as well as other possible sources for correlations. In other words it is important to understand the non-critical background. This can only be done within well understood transport simulations.

In this study we will employ the UrQMD transport model \cite{ppnp41255,JPG251859} with (mean-field mode, UrQMD/M) and without the hadronic potentials (cascade mode, UrQMD/C) to generate a large number of events for nuclear collisions at different bombarding energies. From these events the second order SFMs will be calculated. Similar to what was done in Ref.\cite{NA49SFM}, a correlation-free background can be generated from mixed events, where particles are taken from different events randomly. Any critical behavior is expected to be encoded in the approximate correlator, which can be estimated by the difference between the SSFMs of the real events $F_2^{(d)}(M)$ and the associated mixed events $F_2^{(m)}(M)$:
\begin{equation}
\Delta F_2(M)=F_2^{(d)}(M)-F_2^{(m)}(M).
\label{eq:estimator}
\end{equation}
The power law scaling of the correlator $\Delta F_2(M)$ can be extracted by a fit to the $M^2$ dependence of the critical component ($\Delta F_2(M) \sim (M^{2})^{\phi_{2}}$) for $M \gg 1$ \cite{NA49SFM}, which will then be investigated and compared to the expected critical exponents.

Using the mean-field mode of the UrQMD model \cite{plb659525,plb623395,JPG36015111,mpla27,scpma59632001,mplalpc}, density-dependent potentials for both formed hadron and pre-formed hadron from string fragmentation are treated in a similar way. The density dependent potential reads
\begin{equation}
  U=\alpha\left(\frac{\rho_{h}}{\rho_{0}}\right)+\beta\left(\frac{\rho_{h}}{\rho_{0}}\right)^{\gamma},
\end{equation}
where $\alpha$, $\beta$ and $\gamma$ are parameters. In this work, a soft equation of state is adopted with $\alpha=$-110.49 MeV, $\beta$=182.014 MeV, $\gamma$= 1.17. Here, $\rho_{0}=0.16$ fm$^{-3}$ is the nuclear matter saturation density, and $\rho_{h}=\sum_{i\neq j}c_{i}c_{j}\rho_{ij}$ is the hadronic density, which has contributions from formed and pre-formed baryons with $c_{i,j}=1$, and from pre-formed mesons with a reduction factor $c_{i,j}=2/3$ based on the light-quark counting rules.

The momentum-dependent term of the hadronic potentials reads as:
\begin{equation}
  U_{md}=\sum_{k=1,2}\frac{t_{md}^{k}}{\rho_{0}}\int d\textbf{p}_{j}\frac{f(\textbf{r}_{i},\textbf{p}_{j})}{1+[(\textbf{p}_{i}-\textbf{p}_{j})/a_{md}^{k}]^{2}},
\end{equation}
where $t_{md}$ and $a_{md}$ are parameters. A detailed description on how the model parameters be fixed can be found in Ref. \cite{prc72064908}. For simplicity, the momentum-dependent term is only considered for formed baryons, and a two-body Coulomb potential is included for formed charged particles only.

Furthermore, relativistic effects on the relative distance $\mathbf{r}_{ij}$ and relative momentum $\mathbf{p}_{ij}$ between two particles $i$ and $j$ are taken into account:
\begin{eqnarray}
  \tilde{\mathbf{r}}_{ij}^{2} &=& \mathbf{r}_{ij}^{2}+\gamma_{ij}^{2}(\mathbf{r}_{ij}\cdot \mathbf{\beta}_{ij})^{2}, \\
  \tilde{\mathbf{p}}_{ij}^{2} &=& \mathbf{p}_{ij}^{2}-(E_{i}-E_{j})^{2}+\gamma_{ij}^{2}\left(\frac{m_{i}^{2}-m_{j}^{2}}{E_{i}+E_{j}} \right)^{2},
\end{eqnarray}
where the velocity factor $\beta_{ij}=(\mathbf{p}_{i}+\mathbf{p}_{j})/(E_{i}+E_{j}$) and the $\gamma_{ij}=1/\sqrt{1-\mathbf{\beta}_{ij}^{2}}$. Additionally, $m_{j}/E_{j}$, a covariance-related reduction factor for potentials in the Hamiltonian used as in the simplified version of the relativistic quantum molecular dynamics (RQMD/S) model \cite{prc101034915,prc72064908,PTP96263,EPJA5418,prc97064913}.

\section{results}
\label{results}

In the following we will show the results of the proton intermittency analysis for $^{40}$Ar+$^{45}$Sc, $^{131}$Xe+$^{139}$La collisions with centralities 0-5\%, 5-10\% and 10-15\% and central $^{197}$Au+$^{197}$Au collisions (0-10\%) (the centrality is defined by impact parameter distribution $c=(b/b_{max})^{2}$ of the UrQMD model). The analysis is done for protons (and neutrons) with transverse momenta of $|p_{x,y}|\leqslant1.5$ GeV/$c$, in the mid-rapidity region, i.e. with rapidity $|y|\leq0.75$. To accumulate enough statistics for the analysis, more than 600K events are simulated for each scenario.

\begin{figure}[tb]\centering
\includegraphics[width=0.5\textwidth]{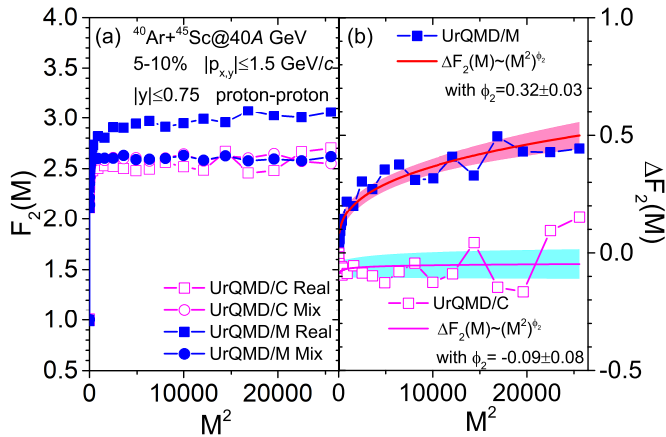}
\caption{(Color online) Left plot: The second scaled factorial moment ($F_2(M)$) of the proton density in transverse momentum space $|p_{x,y}|\leqslant1.5$ GeV/$c$ at mid-rapidity ($|y|\leqslant0.75$) for the 5-10\% central $^{40}$Ar+$^{45}$Sc collisions at 40$A$ GeV obtained from UrQMD model with (UrQMD/M, solid symbols) and without (UrQMD/C, open symbols) hadronic potentials. The squares and circles represent the $F_2(M)$ of the real and mixed events, respectively. Right plot: The estimated correlator $\Delta F_{2}M$ of protons for the corresponding system, shaded bands are 95\% confidence intervals around the fitted lines.}
\label{fig1}
\end{figure}

Fig.\ref{fig1} shows the resulting second scaled factorial moment $F_2(M)$ and the estimator of the correlator $\Delta F_2(M)$ for 5-10\% central $^{40}$Ar+$^{45}$Sc collisions at 40$A$ GeV calculated within UrQMD model with and without hadronic potentials. In panel (a), the intermittency effect, i.e. the difference in $F_{2}(M)$ between real events and mixed events, shows up for $M^2\gtrsim100$.
Therefore, to determine the intermittency index, a fit of $\Delta F_{2}(M)$ using Eq. \ref{Eq:PowerLaw} was used for $M^2>100$. The resulting curve is shown in panel (b) of Fig.\ref{fig1}.
For the UrQMD cascade mode, the results for the moments of the real and mixed events are almost identical, and the values of $\Delta F_2(M^2)$ fluctuate around zero, implying no effect. In the mean-field mode, the $F_2(M)$ of real events rises significantly above those of mixed events, the extracted intermittency index is $\phi_{2}=0.32\pm0.03$. This value of $\phi_2$ is similar to the one found in the NA61 preliminary analysis \cite{NA61}.

\begin{figure}[tb]\centering
\includegraphics[width=0.5\textwidth]{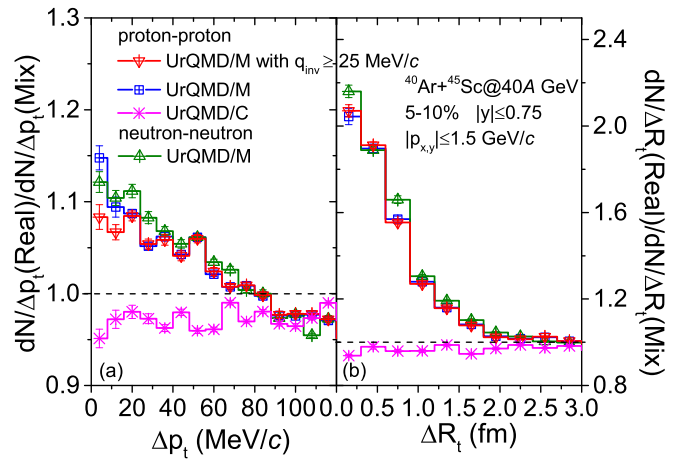}
\caption{(Color online) The correlations function of proton/neutron pairs (ratio of pairs from real and mixed events) at mid-rapidity for the 5-10\% central $^{40}$Ar+$^{45}$Sc collisions at 40$A$ GeV, left panel is for $\Delta p_t$ and right panel is for $\Delta R_t$. The errors are the standard deviation.}
\label{delptpt}
\end{figure}

To understand what causes the non-vanishing intermittency index in the model simulation, we further investigate the baryon pair correlations in momentum and coordinate space. Fig.\ref{delptpt} shows the distribution of $\Delta p_t$ and $\Delta R_t$, over the mixed background, for proton and neutron pairs, in the same simulations. Here, the relative transverse momentum $\Delta p_t$ and the relative radial distance $\Delta R_t$ are defined as:

\begin{eqnarray}
  \Delta p_t &=& \frac{1}{2}\sqrt{(p_{x1}-p_{x2})^2 + (p_{y1}-p_{y2})^2}, \\
  \Delta R_t &=& \frac{1}{2}\sqrt{(r_{x1}-r_{x2})^2 + (r_{y1}-r_{y2})^2}.
\end{eqnarray}

In Fig.\ref{delptpt}, the baseline scenario of the cascade simulation (magenta line with cross symbols) does not show any structure in the relative transverse momentum and coordinate difference distributions as compared to the mixed background. Since in the cascade mode, the hadrons interact with each only through binary scattering according to a geometrical interpretation of elastic and inelastic cross sections. The ratio real/mixed is only slightly below unit due to conservation law effects. On the other hand, the $\Delta p_t$ and $\Delta R_t$ distributions in the simulation with potentials  show a clear increase for small relative transverse momentum $\Delta p_t\lesssim60$ MeV/$c$ and $\Delta R_t \lesssim 1.5$ fm.
To further interpret the observed increase in the $\Delta p_t$ and $\Delta R_t$ distributions, we found that with the consideration of the hadronic potentials, the real collision case has 0.029 proton pairs per event more than that of mixed collisions case at $\Delta p_{t}<$ 64 MeV/$c$. This is true for both, proton (blue line with squares) and neutron pairs (olive line with up triangles), indicating that the Coulomb interaction is less important for this distribution, and the effect of the Coulomb interaction of formed charged particles on this distribution is weaker than that of hadronic potentials. In addition, applying a cut (red line with down triangles), similar to that of the NA61 analysis \cite{NA49SFM} in the relative invariant four-momentum $q_{inv}\geqslant$ 25 MeV/$c$ only has a small effect on the distributions ($\Delta p_{t}\lesssim$20 MeV/$c$).

\begin{figure}[tb]\centering
\includegraphics[width=0.5\textwidth]{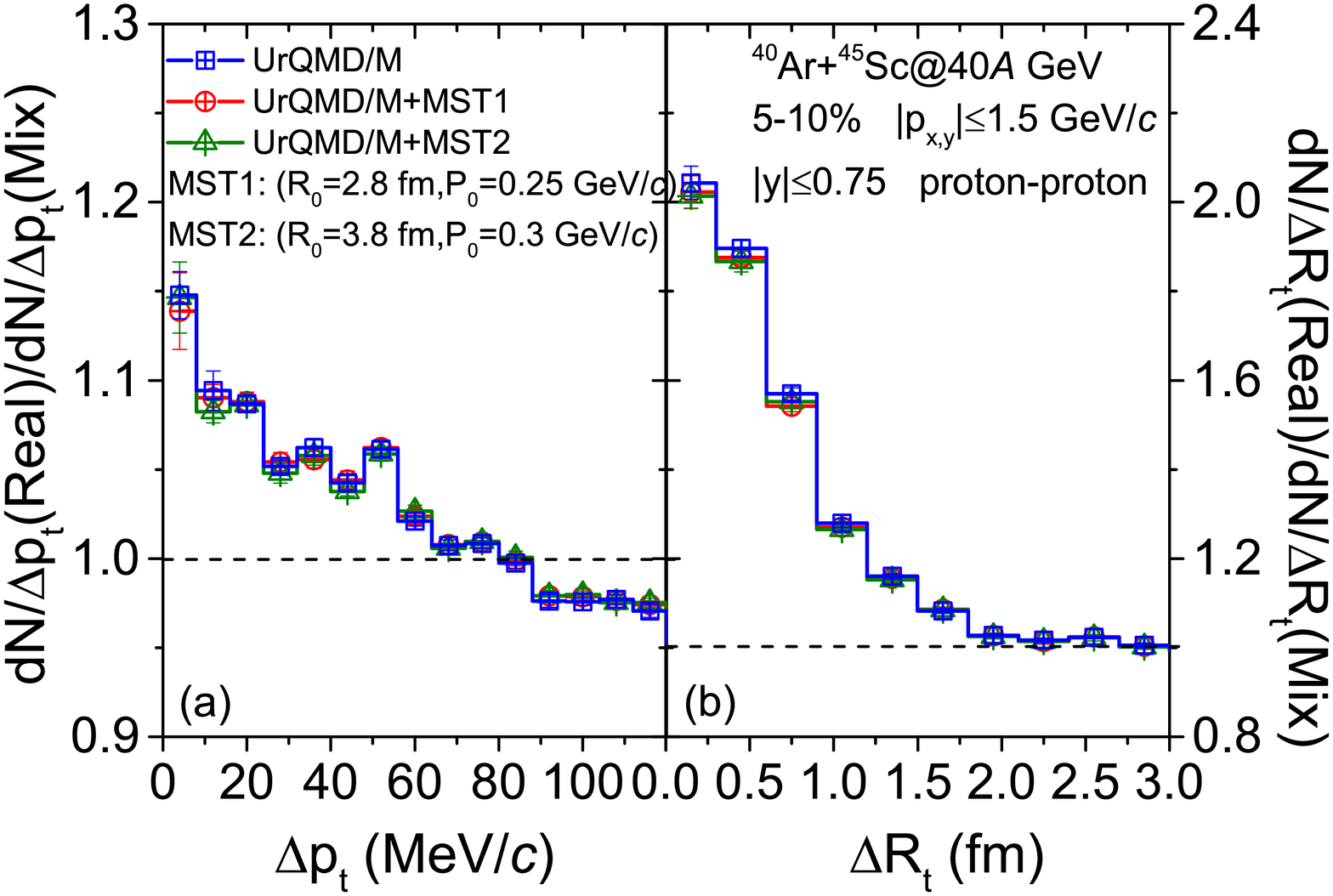}
\caption{(Color online) Similar as Fig.\ref{delptpt}, but for the proton-proton $\Delta p_t$, $\Delta R_t$ distributions with (red line with circles and olive line with triangles) and without (blue line with squares) the traditional coalescence afterburner (Minimum Spanning Tree, MST). The errors are the standard deviation.}
\label{delptptf14}
\end{figure}

Furthermore, to estimate the effect of the coalescence afterburner on these distributions we employ a minimum spanning tree (MST) algorithm for phase space coalescence. Fig.\ref{delptptf14} shows the $\Delta p_t$ and $\Delta R_t$ distributions of proton pairs from UrQMD/M mode with (UrQMD/M+MST) and without (UrQMD/M, blue lines) the coalescence afterburner. The phase-space coalescence afterburner (MST) used the parameter set of ($R_{0}$, $P_{0}$)=(2.8 fm, 0.25 GeV/$c$ and 3.8 fm, 0.3 GeV/$c$), indicated by red lines and olive lines respectively, for the relative momenta and distances \cite{scpma2016}) in the final freeze-out stage as conditions for cluster formation. In the MST algorithm, nonphysical clusters (such as di-neutrons or di-protons) are eliminated by breaking them into free nucleons. The available data of deuteron and Helium-3 productions can be described fairly well by the UrQMD model with the above empirical coalescence parameters\cite{scpma2016,Sombun:2018yqh}. By adjusting the parameter of MST, one can found that the coalescence afterburner has no effect on the proton-proton correlations. This means that the observed correlations in the low $\Delta p_t$ and $\Delta R_t$ can be readily understood as result of the attractive potentials between the nucleons. This attractive interaction eventually also is the reason for nuclear fragment formation, e.g. for the formation of bound pronto-neutron systems, the deuteron.

\subsection{Centrality dependence}

\begin{figure}[tb]\centering
\includegraphics[width=0.5\textwidth]{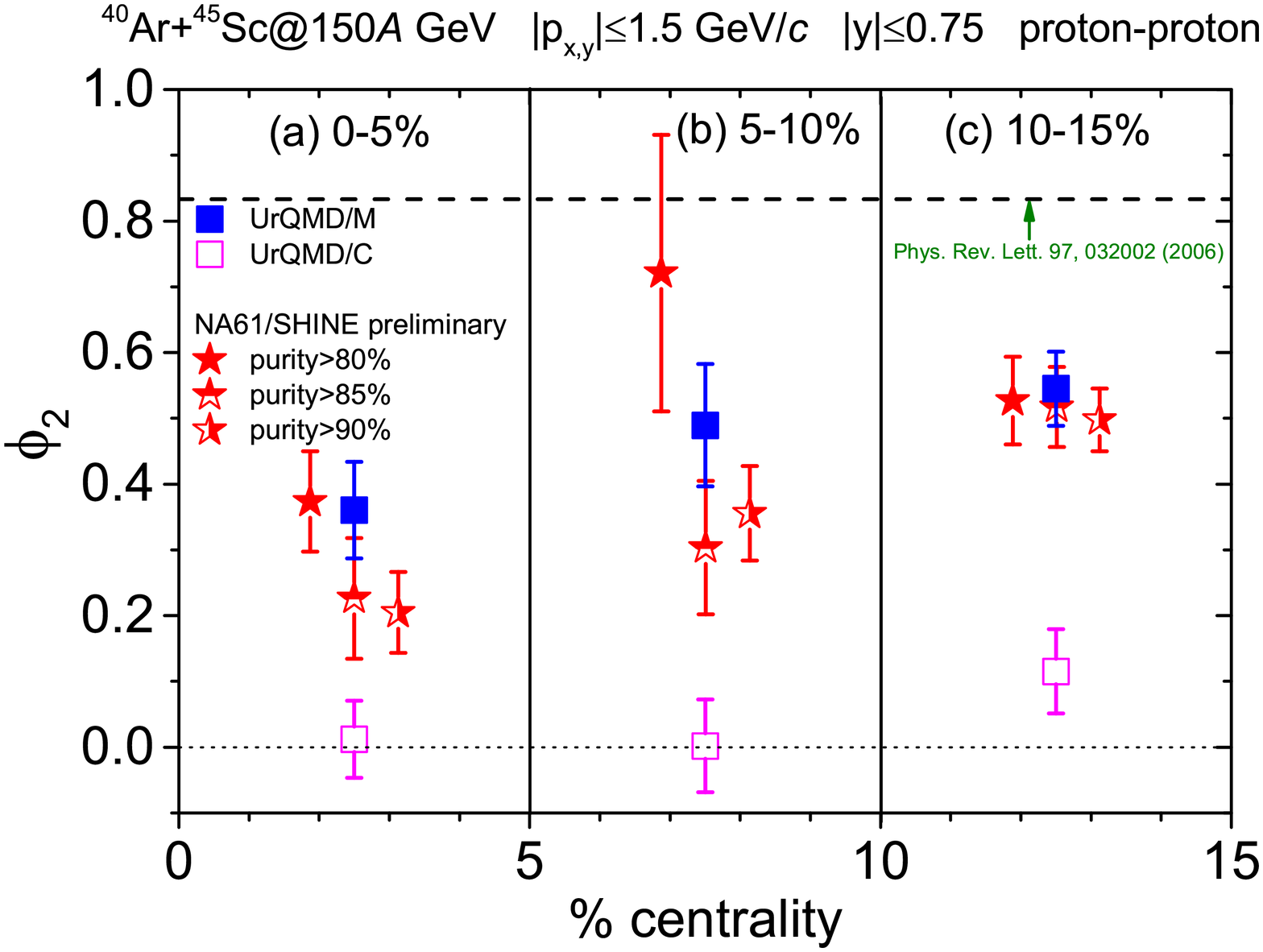}
\caption{(Color online) The second-order intermittency index as a function of centrality in $^{40}$Ar+$^{45}$Sc collisions at 150$A$ GeV, at 0-5\% (left plot), 5-10\% (middle plot) and 10-15\% (right plot) centrality range. The  NA61/SHINE preliminary data are taken from Ref. \cite{NA61}.}
\label{figArScC}
\end{figure}

Next we study the centrality dependence of intermittent fluctuations in the transverse plane. Fig.\ref{figArScC} depicts the centrality dependence of the second-order intermittency index in $^{40}$Ar+$^{45}$Sc collisions. Panels (a), (b) and (c) are for 0-5\%, 5-10\% and 10-15\% collisions, respectively. The calculated results from UrQMD/C (open squares) and UrQMD/M (solid squares) are shown together with the experimental data from NA61/SHINE (stars) with different proton purities \cite{NA61} for comparison. In experiment, the profile of SSFMs is affected by the proton purity selection, thus a full scan in proton purity, at thresholds of 80\%, 85\%, and 90\% are shown. The black dash line presents the expected $\phi_{2}$ for a second-order phase transition calculated based on the effective action belonging to 3D Ising model \cite{Antoniou2006}. The intermittency index calculated by the UrQMD/C model is essentially zero. Further, with the inclusion of hadronic potentials, it is found that the intermittency index is increased and comparable to the NA61/SHINE data in different centrality bins. In addition, it can be seen that $\phi_{2}$ slightly increases with increasing impact parameter, which is also consistent with the findings in the experiment \cite{Davisapb}. This result is due to the decreasing size of the system which goes along with a shortened hadronic freeze out phase \cite{npa20201003,prc73044905,prc90054907}.

\subsection{Energy and system dependence of $\phi_{2}$}

\begin{figure}[t]\centering
\includegraphics[width=0.5\textwidth]{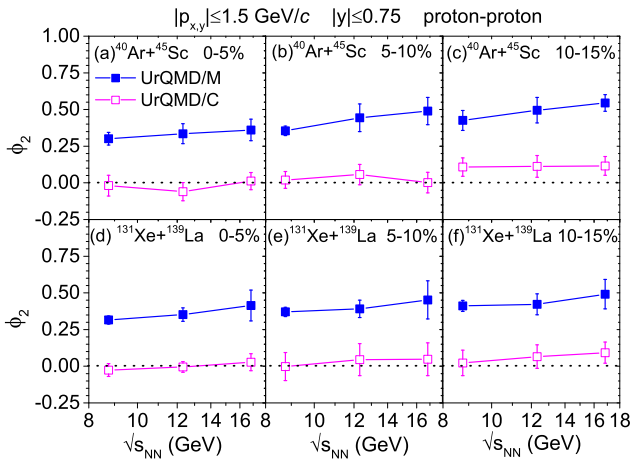}
\caption{(Color online) Excitation functions of the second-order intermittency index in the 0-5\% (panels a and d), 5-10\% (panels b and e), 10-15\% (panels c and f) central $^{40}$Ar+$^{45}$Sc (top plots) and $^{131}$Xe+$^{139}$La (bottom plots) collisions calculated within UrQMD model.}
\label{figArScXeLaE}
\end{figure}

Finally, an energy scan of the intermittency in collisions of $^{40}$Ar+$^{45}$Sc, $^{131}$Xe+$^{139}$La and $^{197}$Au+$^{197}$Au is presented. Fig.\ref{figArScXeLaE} summaries the calculated energy excitation function of the second-order intermittency index. Panels (a), (b) and (c), respectively, are for $^{40}$Ar+$^{45}$Sc at 0-5\%, 5-10\% and 10-15\%. And panels (d), (e) and (f) are for $^{131}$Xe+$^{139}$La collisions. We observe that the $\phi_{2}$ calculated with the UrQMD/C data (open squares) is essentially zero no matter the centrality or energy, and no intermittency effect is observed. In the mean-field mode, the second-order intermittency index (solid squares) is slightly increasing with the energy and/or centrality.

\begin{figure}[t]\centering
\includegraphics[width=0.5\textwidth]{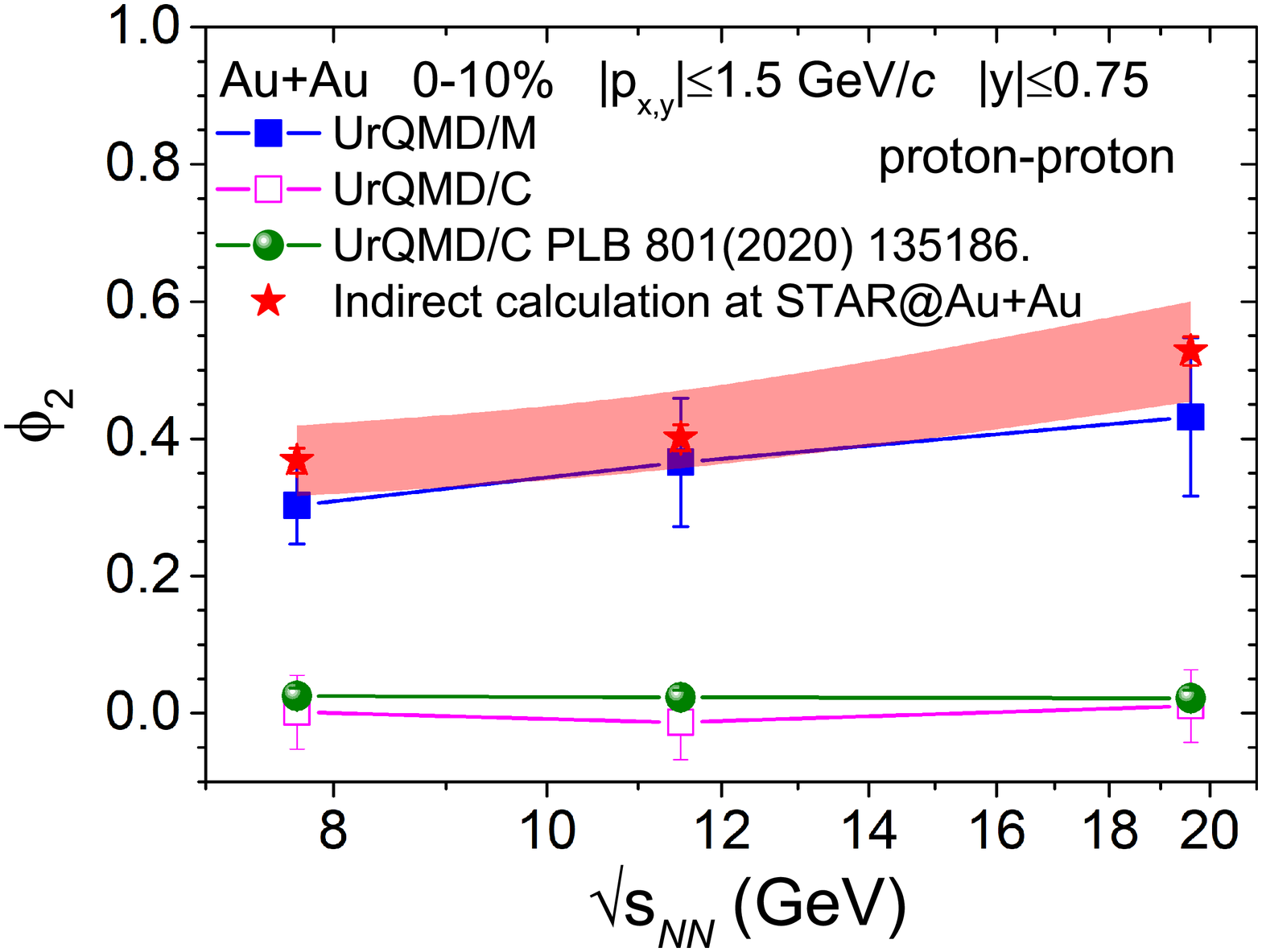}
\caption{(Color online) Energy dependence of the second-order intermittency index in $^{197}$Au+$^{197}$Au collisions. The data which represented by red stars are taken from Ref. \cite{plb801135186} for Au+Au collisions.}
\label{figPbE}
\end{figure}

Fig.\ref{figPbE} shows the excitation function of $\phi_{2}$ from the UrQMD model for 0-10\% central Au+Au collisions at several different energies.
Here, we present results for Au+Au systems as they are studied in the beam energy scan of the STAR experiment. Again, we observe an increase of the intermittency index as function of beam energy. The UrQMD results in Fig.\ref{figPbE} are also compared to conjectured values of $\phi_2$ which are based on a theoretical interpretation of STAR data \cite{plb801135186,arxiv2002}. The data (red stars) from Ref. \cite{plb801135186} are obtained by a mapping of the
$N_{\text{triton}}\cdot N_{\text{proton}}/N_{\text{deuteron}}^{2}$ 
ratio on the neutron number fluctuations in most central (0-10\%) Au + Au collisions \cite{arxiv2002}. These indirectly reconstructed fluctuations are then again mapped onto the obtained relation between the relative density fluctuation of baryons $\Delta$n and $\phi_{2}$. The shadowed band corresponds to the systematic errors from the experimental data. Even though the magnitude and energy dependence of the indirectly calculated data is similar to the model results 
\cite{plb801135186} one should be cautious with this comparison due to the indirect nature of how $\phi_2$ is extracted. A future direct measurement would allow a much more direct comparison with data.

\section{summary}
\label{summary}
Investigations on the second order scaled factorial moments of proton number fluctuations in transverse momentum space of HICs, within the UrQMD transport model, have been presented.
It was found that the inclusion of hadronic mean-field potentials introduce a non-zero intermittency index $\phi_2$ which is consistent with the reported data of the NA49/NA61 experiment.

The energy dependence ($40A$, $80A$, $150A$ GeV), system size ($^{40}Ar$+$^{45}$Sc, $^{131}$Xe+$^{139}$La, $^{197}$Au+$^{197}$Au) and centrality (0-5\%, 5-10\%, 10-15\%) dependence is studied in the UrQMD model with and without hadronic potentials. A clear dependence of beam energy and centrality is observed. In the present study the increase of the SFMs is due to an enhancement of proton pairs (approximately 0.029 pairs per event for the case of 5-10\% central Ar+Sc collisions at 40$A$ GeV) with small relative momenta $\Delta p_{t}\lesssim60$ MeV/$c$ due to attractive nuclear forces. With a further consideration of the traditional coalescence afterburner, the observed correlations of protons were found to not be influenced by the coalescence parameters. More theoretical and experimental studies on the effect of both, hadronic interactions, the stiffness of the equation of state, and the final-state interactions are required to shed light on the intermittency in heavy ion collisions, especially for lower collision energies.

\begin{acknowledgments}
The authors acknowledge support by the computing server C3S2 in Huzhou University. The work is supported in part by the National Natural Science Foundation of China (Nos. 11875125, U2032145, and 12047568), the National Key Research and Development Program of China under Grant No. 2020YFE0202002, and the ``Ten Thousand Talent Program" of Zhejiang province. JS thanks the Samson AG and the BMBF through the ErUM-data project for funding.
\end{acknowledgments}

\end{document}